\documentclass[aps,pra,reprint,groupedaddress]{revtex4-1}
\usepackage[UKenglish]{babel}
\usepackage[]{graphicx, amsfonts, amsmath, amssymb, amstext, latexsym, float, color, hyperref}
\newcommand{\be}{\begin{equation}}
\newcommand{\ee}{\end{equation}}

\begin{document}

\title{Practical Quantum Repeaters with Parametric Down-Conversion Sources}

\author{Hari Krovi, Saikat Guha, Zachary Dutton}
\affiliation{Quantum Information Processing group, Raytheon BBN Technologies, 10 Moulton Street, Cambridge, MA USA 02138}
\author{Joshua A. Slater$^\dag$, Christoph Simon, Wolfgang Tittel}
\affiliation{Institute for Quantum Science and Technology, Department of Physics and Astronomy, University of Calgary, Alberta T2N 1N4, Canada \\
$^\dag$Present address: Vienna center for quantum science and technology (VCQ), Faculty of Physics, University of Vienna, 1090 Vienna, Austria}


\begin{abstract}
Conventional wisdom suggests that realistic quantum repeaters will require quasi-deterministic sources of entangled photon pairs. In contrast, we here study a quantum repeater architecture that uses simple parametric down-conversion sources, as well as frequency-multiplexed multimode quantum memories and photon-number resolving detectors. We show that this approach can significantly extend quantum communication distances compared to direct transmission. This shows that important trade-offs are possible between the different components of quantum repeater architectures.
\end{abstract}


\maketitle

\section{Introduction}

The distribution of quantum states over long distances is essential for applications such as quantum key distribution \cite{QKD} and a future ``quantum internet'' \cite{kimble}. The distances that are accessible by direct transmission are limited by photon loss. Some form of quantum repeater \cite{briegel} architecture will be required to overcome this barrier. On the one hand, approaches based on quantum error correction \cite{QEC} or satellite links \cite{boone} promise quantum communication over global distances in the long term, but have significant resource requirements. On the other hand, there is also a lot of interest in simpler approaches, where the focus is on significantly outperforming direct transmission in the short or medium term \cite{sangouard}.

The first concrete proposal for a quantum communication architecture with relatively modest resource requirements was the well-known DLCZ protocol \cite{DLCZ}. It is based on atomic ensembles that serve as photon sources and quantum memories. This proposal stimulated a lot of experimental work \cite{DLCZexp}, but it was soon recognized that the achievable repeater rates were still too low. A potential solution was put forward in \cite{Simon07}, which proposed to implement a multiplexed version of the DLCZ protocol combining parametric down-conversion sources (which are comparatively simple to implement) and multimode quantum memories. Such memories are now being developed very intensively, in particular in rare-earth doped crystals \cite{tittelreview,afc}.

Unfortunately, the protocol of \cite{Simon07} (just as the DLCZ protocol) relies on single-photon interference to create entanglement in the elementary repeater links, and thus requires interferometric stability over long distances, which is a major practical challenge. This difficulty can be avoided by designing repeater protocols where the elementary entanglement creation is based on two-photon interference \cite{Sangouard08,Sinclair14}.

However, in the present context, relying on two-photon interference also means relying on simultaneous single photon pair emissions from two different sources, so that one photon from each pair can interfere. It is then challenging to work with parametric down-conversion sources because they can always emit multiple pairs, which typically causes errors. Some of these errors can be eliminated by working with small emission probabilities, but this has a large negative impact on the achievable rates. In prior work \cite{guha}, we showed that even a small amount of multiple pair probability can be detrimental when single photon detectors are used. This would preclude the use of parametric down-converters along with single photon detectors since the multiple pair probability is much higher in down-converters.

Past proposals therefore focused on quasi-deterministic sources of entangled photon pairs, which can in principle be realized using individual emitters such as atoms or quantum dots \cite{directpairsources}, 
or by combining parametric down-conversion with strong nonlinearities \cite{zeno}. While many of these approaches seem promising in the longer term, they all pose significant practical challenges in the short and medium term. Recently in \cite{Takeoka}, an analysis of entanglement swapping using down-converter sources and single photon detectors has been done using Wigner functions. In \cite{Khalique}, numerical analysis of a relay architecture with down-converter sources was done for up to three links and it was found that the rates came close to the TGW bound.

Here we adopt a different approach. We reconsider the issue of realizing quantum repeaters with down-conversion sources and two-photon interference, and we show that this approach can in fact lead to impressive quantum repeater performance (with a scaling that clearly beats the TGW bound), provided that there are two additional elements, namely highly multi-mode quantum memories and photon-number plus spectrally resolving detectors. Multi-mode memories help to compensate for the requirement of working with low pair emission probability, and photon-number resolving detectors make it possible to identify and greatly remove the remaining errors due to multi-pair emissions. Both highly multi-mode memories and photon-number resolving detectors are under very active development at this point. The main conclusion from our study is therefore that truly practical quantum repeaters may be within reach. Our results also show that in designing quantum repeater architectures there are interesting trade-offs between the performance and capabilities of the different components, i.e. in the present case, pair sources, memories, and detectors.

\section{Repeater Architecture}

\subsection{Description of the Scheme}

The architecture that we consider is similar to that proposed in \cite{Sinclair14} and analyzed in detail in \cite{guha}, but with the key difference that we consider parametric down-conversion sources and photon-number resolving detectors, instead of close to ideal pair sources and non-number resolving detectors.

\begin{figure}
\centering
\includegraphics[width=\columnwidth]{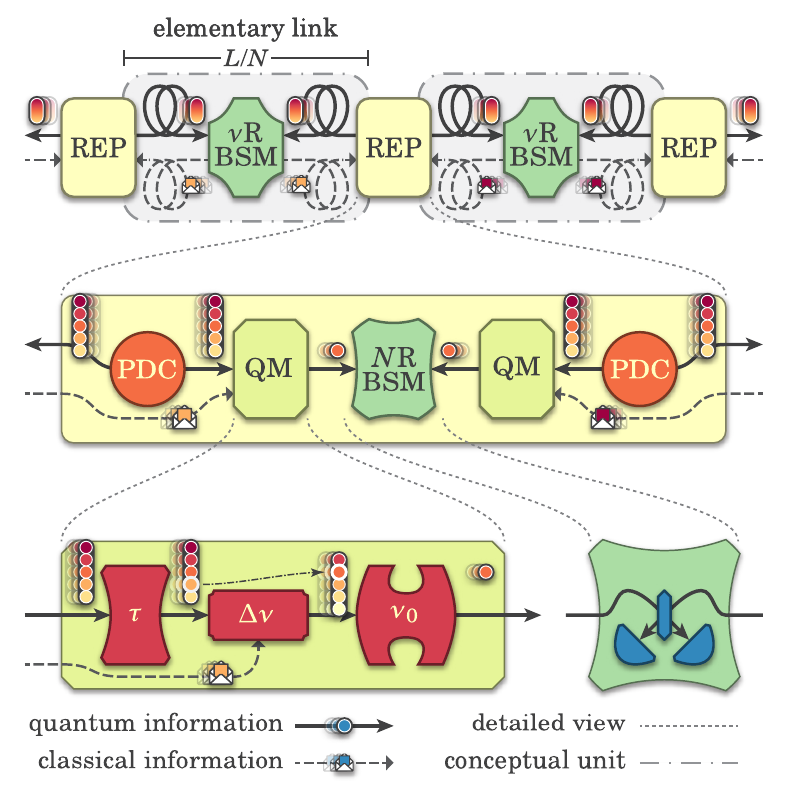}
\caption{Schematic of quantum repeater architecture~\cite{Sinclair14}. Top Level: The channel is divided into $N$ elementary links, each with a spectrally-resolving BSM ($\nu$RBSM) at its center, and with Repeater Nodes (REP) connecting each pair of links.  Middle Level: Detailed view of Repeater Node.  Each REP contains two PDC sources of frequency-multiplexed bi-partite entanglement, two quantum memory units (QM) and a number-resolving BSM ($N$RBSM).  Bottom Level: Detailed view of QM and $N$RBSM.  Each QM unit contains a multi-mode atomic quantum storage device ($\tau$), a frequency-shifting unit ($\Delta\nu$) and a frequency filter ($\nu_0$), while each $N$RBSM contains a simple linear-optics circuit followed by two single-photon detectors.  See main text for a detailed description of the functioning of the architecture}
\label{fig:architecture}
\end{figure}

The architecture~ is depicted schematically in Fig.~\ref{fig:architecture}. The total distance between Alice to Bob is divided into $N$ elementary links. At either side of each elementary link is a repeater node (REP) containing a parametric down-conversion source (PDC), which sends one half of an entangled state to the center of the elementary link through a fiber. At every time step entangled states are created in a large number of modes $M$. We envision $M$ of the order $10^6$, which can be achieved by using many distinct frequency modes, or by combining spectral and spatial multiplexing (e.g. 100 spatial modes), see below. The entanglement in each mode can be in the polarization or temporal/time-bin degree of freedom, as our discussion applies equally to either encoding. At the center of each elementary link, a frequency-resolving linear-optic Bell-state measurement ($\nu$RBSM) is performed on the state comprising of the halves of the two entangled pairs coming from each side. This creates an entangled state across the elementary link by entanglement swapping. The $\nu$RBSM consists of a simple linear optical circuit ~\cite{Bra95} followed by a spectrally-resolved detector array \cite{Allman}, which can detect across a range of frequencies. Note that the efficiency of a linear optical BSM can at most be 50\% for each mode \cite{Lutkenhaus}. The other half of each of the entangled states produced by the sources is locally loaded into a multi-mode atomic quantum memory (QM)~\cite{Sinclair14}. 

For realistic link lengths, most of the photons produced by the sources will be lost in transit to the centre of an elementary link. However, if one chooses a large number of modes $M$, then a successful swap for at least one of them occurs with a high probability. At this point, this frequency information (i.e. the successful frequencies) is transmitted to the memories on either side of the elementary link (dashed lines and envelopes in Fig.~\ref{fig:architecture}). Now, each repeater node (REP) receives a pair of which-frequency information from the two $\nu$RBSMs on either side. It uses this information to translate the frequency of the modes (device labelled $\Delta\nu$) on either side to a pre-determined common frequency and filter away all other modes (device labelled $\nu_0$)~\cite{Sinclair14}, and thereafter performs a linear-optic BSM (with number-resolving detectors, see below) to do entanglement swapping (NRBSM). If this BSM is successful, the states across two elementary links are connected to create an entangled state across both of them. This process is continued until we obtain an entangled state across Alice and Bob. Alice and Bob can now use this entanglement for tasks such as quantum key distribution or quantum teleportation. For the case of quantum key distribution, secret key rates can be determined following the approach of \cite{guha}.

\subsection{Parametric Down-Conversion Sources}

We now describe the entangled state generated by the parametric down-conversion sources in more detail. For each of the $M$ (frequency) modes discussed above, each source generates a multi-photon entangled state of the form $|\psi\rangle=e^{-iHt}|vac\rangle$, where
\begin{equation}
H=ig(a^{\dagger}_0 b^{\dagger}_1 - a^{\dagger}_1 b^{\dagger}_0) + h.c.,
\end{equation}
where the coupling constant $g$ is proportional to the pump laser amplitude and nonlinear coefficient of the crystal, the creation operators $a_i$ and $b_i$ refer to the two ``halves'' of the entangled state discussed above, and the index $i=0,1$ refers to a mode of the degree of freedom in which the entanglement is prepared, i.e. either polarization or time bins. Each ``mode'' in our above terminology therefore really corresponds to four physical modes $a_0, a_1, b_0, b_1$. One can show that \cite{multiphotonPDC}
\begin{equation}\label{eq:PDC_state}
|\psi\rangle=\sum_{n=0}^{\infty} \frac{\sqrt{n+1}\tanh^n gt}{\cosh^2 gt} |\psi_n\rangle,
\end{equation}
with
\begin{eqnarray}\label{eq:PDC_terms}
|\psi_n\rangle=\frac{1}{n!\sqrt{n+1}} (a^{\dagger}_0 b^{\dagger}_1 - a^{\dagger}_1 b^{\dagger}_0)^n |vac\rangle= \nonumber\\ \frac{1}{\sqrt{n+1}}\sum_{m=0}^n (-1)^m |n-m,m;m,n-m\rangle,
\end{eqnarray}
where $|n-m,m;m,n-m\rangle$ signifies a state with $n-m$ photons in mode $a_0$, $m$ photons in mode $a_1$, $m$ photons in mode $b_0$ and $n-m$ photons in mode $b_1$.

In the context of the quantum repeater protocol described above, only the term with $n=1$, corresponding to the emission of a single entangled photon pair, is desired. The case $n=0$ means that no photons were emitted at all, whereas the terms with $n \geq 2$ correspond to multi-pair emissions, which a priori introduce errors. We now discuss how these errors can be greatly suppressed using photon-number resolving detectors.

\subsection{Suppression of Multi-photon Errors using photon-number resolving detectors}

Ref. \cite{guha} analyzed multi-pair errors in the context of the repeater protocol of Ref. \cite{Sinclair14} (i.e. for much more ideal sources than parametric down-conversion) and found that they severely limit its performance. This analysis was done for ordinary single-photon detectors, which do not count the number of photons. However, photon-number resolving (PNR) detectors are being developed and have reached impressive performance levels \cite{lita}. We now show that the use of such detectors allows one to greatly alleviate the problems associated with multi-pair emission by down-conversion sources. In fact, we show that ideal PNR detectors, together with ideal quantum memories, would allow one to eliminate the associated multi-photon terms completely in this repeater architecture.

The improved performance with the use of PNR detectors can best be understood by considering perfect PNR detectors and quantum memories (i.e.~$100\%$ efficiency and no noise). We argue that by post-selecting the outcomes corresponding to single pair terms at the repeaters as well as by Alice and Bob, one can completely eliminate the multi-photon errors. At the center of each elementary link one can use either ordinary single photon detectors or PNR detectors. With this setup, first, one can post-select the outcomes corresponding to single pair terms at Alice's and Bob's ends. This means that Alice and Bob wait for a single click in their detectors (which are PNR) and so they know for sure that they have a single photon. This, in turn, means that the entanglement source closest to them has produced the correct state. For simplicity, let us focus on the case when there are two elementary links with a repeater in the center i.e., one elementary link between Alice and the repeater and one between Bob and the repeater. In this case, the center of the elementary link has the right state on one side. If the Bell swap results in two detector clicks, this means that the other side of the elementary link must have had one or more photons. Now reasoning similarly from Bob's side, we arrive at the fact that at the repeater station, the two sources on either side have produced one or more photons. But since the repeater has perfect PNR detectors and we post-select the two photon outcome, we can be sure that the two sources on either side have produced a single pair (i.e., the correct state). Finally, this implies that Alice and Bob share a perfect entangled state. This analysis can be extended to any number of elementary links. This shows that by post-selecting on the single click outcomes by Alice and Bob and the two click outcomes for the repeaters, we can obtain a good entangled state across Alice and Bob. In the presence of memory and detector imperfections such as non-unit efficiency and dark counts, the generated state is no longer perfect, but the fidelity can still be high. In the next section we show what secret key rates should be achievable under realistic conditions with this approach.

\section{Results on Repeater Rates}

\begin{figure}
\centering
\includegraphics[width=\columnwidth]{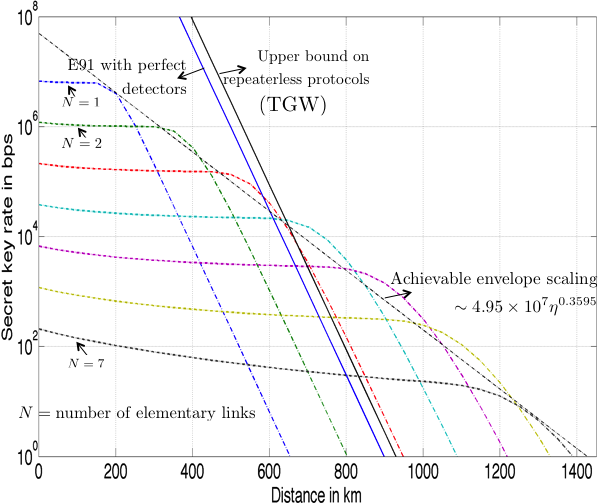}
\caption{Secret key rate vs. distance for SPDC sources with PNR detectors. The parameters used are mean photon number $N_s=0.035$, number of frequency modes of $10^4$, number of spatial modes of $100$, dark click rate = $1$Hz, detector efficiency = $0.95$, memory efficiency = $0.9$, repetition rate = $30$MHz.}
\label{fig:results}
\end{figure}

To quantify the performance of this protocol we utilized our previously developed \cite{guha} MATLAB code which can evaluate entanglement generation and secret key rates across a chain of repeater nodes, accounting for multiple sources of imperfections including detector inefficiency and dark counts, multi-pair emission in the sources, memory lifetime and read/write inefficiency, mode mis-matches in Bell measurements, and photon loss in optical components.   Our analysis propagates the density matrix state present in a single (frequency) mode, initializing with the entangled states created at the sources given in Eq.(\ref{eq:SPDC}) (described in more detail below) and then transforming directly through lossy fiber and then beam-splitters and other components to perform Bell-state measurements at the center of each elementary link. When a state comes upon a detector, a POVM is applied as described below. The POVM properly accounts for the detector inefficiency and dark counts. The remaining (undetected) modes continue to propagate through the remainder of the system. Any instance in which the number or pattern of clicks does not correspond to the ideal case is discarded as an successful attempt and the calculation reveals the probability of a successful attempt for each mode $P_{s0}$. Since we assume large-scale multiplexing $M$  (via frequency and spatial modes) on each elementary link, we then calculate the much higher probability of at least one successful event $P_s(1) =1 - (1-P_{s0})^M$  at a particular elementary link before continuing. We then calculate the propagation through the Bell measurements at each repeater node. When $N$ is a power of two, one can take advantage of symmetry and explicitly calculate a case where the Bell-state measurements are done in a binary tree fashion, with the entanglement distance doubling at each stage. However, our calculation and code are for general cases which do not contain a number of elementary links corresponding to a power of two (see Fig. \ref{fig:results}). Again, the probability of a successful click pattern is calculated and unsuccessful patterns can be discarded. For any successful click pattern, the density matrix can then be used to calculate the total error $Q$ of mis-matched bit values measured at Alice and Bob. The obtainable rate of secret key generation is related to this error rate via $R(Q) = 0.5-h_2(Q)$ where $h_2(x) = -x \log_x(x) - (1-x) \log_2 (1-x)$ is the binary entropy function.  This rate is non-zero when $Q \le 0.1104$. Note that in \cite{guha} we performed this same calculation analytically for the case of perfect entangled photon pair sources and also numerically for the more general source case.

In Fig.~\ref{fig:results}, we plot the secret key rates vs. distance (in log scale) for a number of elementary links from 1 through 7. This is done for SPDC sources with PNR detectors. The mean photon number $N_s$ of the SPDC sources has been chosen so as the optimize the performance. For the parameters chosen, this turns out to be $N_s=0.035$. The other parameters are as shown in the figure. We also plot the achievable rate by repeaterless protocols and also the recent upper bound on the rates of all repeaterless protocols, known as the Takeoka-Guha-Wilde (TGW) bound \cite{TGW}. It can be seen that the envelope over the different numbers of elementary links beats this upper bound for d$>\sim$700 km. The envelope in this case (unlike the case of deterministic sources analyzed in \cite{guha}) is not a straight line in the log scale of the plot. However, as shown in the figure, a straight line fit that is achievable (in the sense that it is below the true envelope) clearly shows that this scaling beats the TGW bound.

{\bf Source modeling:} The initial state can be written as follows.
\begin{align}\label{eq:SPDC}
&|\psi\rangle = \sqrt{p(0)}\,|00,00\rangle + \sqrt{p(1)/2}\,(|10,01\rangle + |01,10\rangle ) \nonumber \\
&+ \sqrt{p(2)/3}\,\left(|20,02\rangle - |11,11\rangle + |02,20\rangle\right)\,.
\end{align}
This is the truncated version of the state in Eq.~\ref{eq:PDC_state} up to two pair terms (or $n=2$ in Eq.~\ref{eq:PDC_state}). The amplitudes become
\begin{align}
&p(0)=1/(N_s+1)\nonumber \\
&p(1)=N_s/(N_s+1)^2\nonumber \\
&p(2)=1-N_s/(N_s+1)^2-1/(N_s+1)\,,
\end{align}
where $N_s=\cosh^2 gt -1$. In contrast, the perfect source (i.e., sources that produce an entangled pair deterministically with no higher photon terms), which is analyzed in detail \cite{guha} is the maximally entangled state
\be
|\psi\rangle = \frac{1}{\sqrt{2}}(|10,01\rangle + |01,10\rangle)\,.
\ee

{\bf Detector modeling:} Detectors are modeled as positive operator valued measurements (POVM) in our simulations. For ideal PNR detectors, the operators with are simply $\{\Pi_0,\Pi_1, \Pi_2\dots\}$, where the $\Pi_n=|n\rangle\langle n|$ signifies the presence of $n$ photons.  Now suppose that the detectors have a sub-unity detection efficiency $\eta$---which may be thought of as arising from a beamsplitter with transmissivity ${\eta}$ just in front of an ideal detector---and independently there may also be a probability $P_d$ for the detector to trigger in the absence of a photon (dark click probability). This means the ``no click'' and ``1-click'' etc., events in the individual detectors correspond to a POVM with outcomes $\{F_0, F_1, F_2,\dots \}$, with
\begin{eqnarray}
F_0 &=& (1-P_d)\Pi_0 + (1-\eta)(1-P_d)\Pi_1  \\
&+& (1-\eta)^2(1-P_d)\Pi_2 \nonumber\\
F_1 &=& P_d\Pi_0 + (\eta + (1-\eta)P_d)\Pi_1 \\
&+& (\eta(1-\eta) + (1-\eta)^2P_d)\Pi_2 \nonumber \\
F_2 &=& \eta^2 \Pi_2\,.
\end{eqnarray}
One can interpret the above expressions as follows. For instance for $F_0$ (the ``no click'' outcome), if there are no actual photons present, one will get this outcome with probability $1-P_d$ i.e., the detector did not have a dark click. On the other hand, if there is a single photon present both it is lost and there is no dark click i.e., with probability $(1-P_d)(1-\eta)$. Finally, it there are two photons are present, both of them are lost and there are no dark clicks i.e., $(1-P_d)(1-\eta)^2$. For single photon detectors (SPD), there are only two POVM outcomes ``click" and the ``no click", which can be labeled $F_0$ and $F_1$. These can be written as from above as
\begin{eqnarray}
F_0 &=& (1-P_d)\Pi_0 + (1-\eta)(1-P_d)\Pi_1  \\
&+& (1-\eta)^2(1-P_d)\Pi_2 \nonumber \\
F_1&=&I-F_0\,,
\end{eqnarray}
where $I$ is the identity operator.


\section{Implementation}

We will now briefly discuss the implementation of our proposed scheme. Parametric down-conversion sources can easily achieve the required bandwidth (of order 300 GHz). One furthermore has to ensure that the sources produce pure states for each frequency mode, which can be achieved by placing the sources inside moderate-finesse cavities, in which the resonances define the spectral modes \cite{PDCcavities}.

Highly efficient (up to 95\%) photon number resolving detectors have already been demonstrated (superconducting transition edge sensors, and superconducting nanowire single photon detectors) \cite{lita,marsili} and intrinsic dark counts in these detectors are extremely low. Dark counts are dominated by black-body radiation and, with appropriate filtering, rates as low as 1 Hz should be achievable \cite{marsili}.
Frequency resolved detection is possible by combining spectrometers with detector arrays. The frequency resolution required here (of order 30 MHz) is in principle feasible with this approach \cite{freqresolution}, and it should also be possible, though challenging, to include spatial resolution \cite{privatecommunication}.

Frequency multiplexed quantum memories can be realized based on rare-earth doped crystals. Optical transitions in these systems can have very large ratios of inhomogeneous (hundreds of GHz) to homogeneous linewidths (kHz), making a great number of frequency channels available \cite{tittelreview}. The largest reported value of inhomogeneous broadening in RE crystals is 300 GHz in Tm:LiNbO$_3$ \cite{TmLNO}. This would allow $10^4$ frequency modes separated by 30 MHz, motivating the values we picked in Figure 2. The proposed repeater architecture does not require readout on demand in time, i.e. it requires only a low-loss delay, followed by appropriate frequency shifts. Such a delay can be realized based on the atomic frequency comb (AFC) memory protocol \cite{afc}. For a delay one needs no control pulses, which makes it easier to achieve high efficiency, and no additional ground state level, which broadens the class of suitable materials. One very promising material is Tm:YGG \cite{thiel}, where storage times of order 1 ms should be possible, corresponding to potential elementary link lengths of order 200 km (as assumed in Figure 2). It should be possible to increase the inhomogeneous broadening of Tm:YGG from its current value of 57 GHz to hundreds of GHz through suitable co-doping with other rare-earth ions, as has been demonstrated with similar materials in the past \cite{codoping}. The memory efficiency can be made very high by using low-finesse cavities \cite{cavities}.

Spatial multiplexing in optical telecommunication is becoming recognized as an important technology to ameliorate the problems of the limited capacity of single-mode fibers, and costs and challenges of adding and/or obtaining additional single-mode fibers. Being developed are techniques to utilize multiple spatial modes in multi-mode fibre cores, multiple cores in a single fiber \cite{sakaguchi}, and combinations of the two \cite{vanuden,mizuno}. Already, up to 36 spatial modes has been demonstrated in the same fiber \cite{mizuno}. These technologies will easily function with quantum communication protocols, including our multiplexing quantum repeater architecture, and allow for significantly more simultaneously, and easily, transmitted modes for multiplexed entangled qubits.

With similar motivations, quantum information researchers have been demonstrating spatially multiplexed entanglement sources and quantum memories, including eight photon-pair sources integrated on the same chip \cite{collins}, and memories capable of storing and retrieving highly multimode spatial structures \cite{heinze}.  The combination of these developing technologies makes 100 spatial modes for spatial multiplexing (as assumed in Figure 2) seem feasible in the near future.

\section{Conclusions and Outlook}

We have shown that it is possible to construct a quantum repeater architecture based on parametric down-conversion sources that clearly outperforms any conceivable repeaterless protocol. The key ingredients for making this possible are highly multi-mode quantum memories and photon-number resolving detectors. Both of these components are currently under very active development and have already reached impressive performance levels. Our results suggest that truly practical quantum repeaters may really be within reach. We would like to emphasize that the proposed architecture is modular, i.e. it can easily accommodate more ideal entangled photon pair sources as they become available. This would either allow to relax the requirements on the memories and detectors, or it would lead to further improvements in performance.
 
\section{Acknowledgments}

This work was supported by Alberta Innovates Technology Futures (AITF), the National Engineering and Research Council of Canada (NSERC), the DARPA Quiness program subaward
contract number SP0020412-PROJ0005188, under prime contract number W31P4Q-13-1-0004.  W.T. is a senior fellow of the Canadian Institute for Advanced Research. We thank Chris Fuchs for useful discussions on modeling, Matt Shaw for useful discussions on detectors and J. Schm\"ole for graphical support.


%

\end{document}